# Compound Hertzian Chain Model for Copper-Carbon Nanocomposites' Absorption Spectrum

Alireza Kokabi, Mehdi Hosseini, Saman Saeedi, Ali Moftakharzadeh, Mohammad Ali Vesaghi, and Mehdi Fardmanesh

*Abstract*—The infrared range optical absorption mechanism of Carbon-Copper composite thin layer coated on the Diamond-Like Carbon (DLC) buffer layer has been investigated. By consideration of weak interactions between copper nanoparticles in their network, optical absorption is modeled using their coherent dipole behavior induced by the electromagnetic radiation. The copper nanoparticles in the bulk of carbon are assumed as a chain of plasmonic dipoles, which have coupling resonance. Considering nearest neighbor interactions for this metallic nanoparticles, surface plasmon resonance frequency ($\omega_0$) and coupled plasmon resonance frequency ($\omega_1$) have been computed. The damping rate versus wavelength is derived which leads to the derivation of the optical absorption spectrum in the term of $\omega_0$ and $\omega_1$. The dependency of the absorption peaks to the particle-size and the particle mean spacing is also investigated. The absorption spectrum is measured for different Cu-C thin films with various Cu particle size and spacing. The experimental results of absorption are compared with the obtained analytical ones.

*Index Terms*—Nanoparticle, Absorption, Nanocomposite, Plasmon resonance

## I. INTRODUCTION

Absorption mechanism of metallic nanoparticles has attracted attentions recently [1]-[10], and a wide range of application has been proposed due to high electromagnetic field energy near the surface of nanoparticles such as Surface Enhanced Raman Spectroscopy [11]. In addition, the effect of molecular absorption on the intensity and position of absorption peak frequency have been studied which led to development of molecular sensors [8]. Noble metalic nanoparticle chains are also applied for detection of the cancerous cell due to their optical absorption [12].

Electron-surface scattering is believed to contribute in the absorption for small particle sizes below 10nm, or for high surface state densities, where chemical interface damping is important [13]. For particle size in the range of nanoparticles, the absorption could not thoroughly be associated to skin effect because the dimensions of the particles are well below the range of skin depth [6]. Especially, in addition, previously published experimental data show that the absorption spectrum of the copper nanoparticle network of Cu-C composite layer does not completely match to the spectrum governed by electron-surface scattering [14].

Another absorption mechanism in the considered structure is attributed to self and mutual interactions of dipoles of nanoparticles with the electromagnetic field. These dipoles are formed by collective oscillation of the electrons, which is known as plasmon. Novel metals such as Cu or Au have considerable absorption in the micrometer range due to surface plasmon resonance and various experiments show that observed coupling between particle dipoles result in absorption frequency peaks, which are red-shifted in compared to that of a single particle [2], [3]. The red-shift effect depends on inter-particle spacing and particle size of the nanoparticles. The red-shift decreases exponentially with increasing particle spacing and the size of nanoparticles. [4] Hence, this spectral characteristic reflects the geometrical properties such as size and spacing of nanoparticles that represent the coupling between dipoles.

Previously proposed models explained the optical absorption of finite one-dimensional particle chains parallel to the nanolayer surface [2]. Brongersma, et al. have modeled the nanoparticle network as a 1D infinite nanoparticle chain to investigate the coupled dispersion relation of electromagnetic wave for this plasmonic system [15]. They have used numerical values of the first and the second resonance frequencies for their proposed equations. They have calculated approximate dispersion relation of the system in the low damping regime. Using their proposed equations, we have also calculated the absorption spectrum of the nanoparticle chain in the high damping regime as a result of that work. Higher precision of spectrum calculation can be obtained by considering the size dependency of resonance frequency for a single nanosphere.

In this work, the metallic nanoparticles in the bulk of the host material are modeled as the chain of the nanospheres, which are arranged in the direction of the electromagnetic radiation. This modeling needs the natural self- resonance frequency as a function of the particle size for different nanoparticle geometries. This size-dependent resonance frequency of metallic nanoparticles is investigated in previous works [5, 7]. We have incorporated the size-dependent natural self-resonance frequency of the single nanosphere to attain a precise model of network of nanoparticles based on the Brongersma's work. These results are compared with the experimental results.

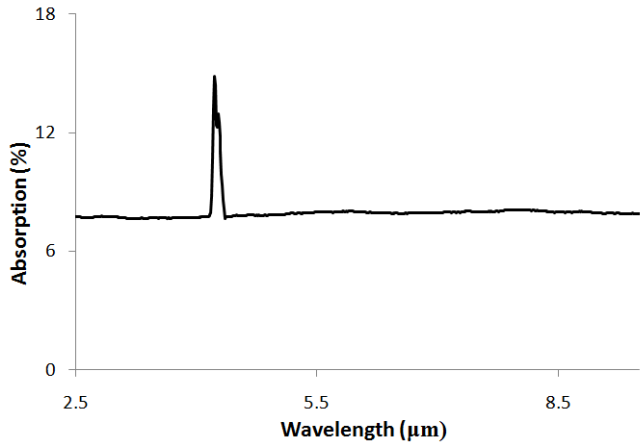
Fig. 1. Absorption spectrum for deposited DLC.

In the absorbing composite layer studied in this work Cu nanoparticles are embedded into a diamond-like carbon (DLC) layer, which has a low absorption. Thus, in this work we neglect the DLC layer absorption compared to that of Cu nano-particles and attribute total absorption of this layer associated to the Cu nanoparticles.

## II. Theoretical Model

The studied nanosphere aggregate caused by radiation absorption is considered monotonous at the surface, which are subject to the perpendicular electromagnetic radiation while the bulk density is different. This assumption is to explain the non-uniformity of nanoparticle network in all dimensions due to deposition methods.

M. L. Brongersma *et. al.* have proposed a model of compound Hertzian chain, which are considered parallel to the electromagnetic radiation propagation [15]. Each of chains consists of nanoparticles which behaves like coupled dipoles. Metallic nanoparticles are assumed as identical spheres with radius *r*, which are uniformly distributed over the chains with a center-to-center distance, *d*.

Equation of motion for charges in term of plasmon-wave solution through a chain of coupled dipoles with consideration of nearest neighbor interactions is as (1). The damping factors like interaction with phonons, electrons, lattice defects and radiation into the far field are covered in these coupled equations [15]

$$-\omega^2 + \omega_0^2 + 2\omega_1^2 \cos(kd)\cosh(\alpha d) = 0$$
$$\omega\Gamma_I + \frac{\omega^3 \Gamma_R}{\omega_0^2} + 2\omega_1^2 \sin(kd)\sinh(\alpha d) = 0 \quad , \quad (1)$$

in which, $\omega$ is the angular frequency of incident radiation, $\omega_0$ is the self plasmonic angular resonance frequency of metallic nanospheres, $\omega_1$ is coupled resonance frequency of two neighboring dipoles, *k* is the wave vector and $\alpha$ is the absorption coefficient. The constants $\Gamma_I$ and $\Gamma_R$ are the electronic relaxation frequencies, which is derived through Mathiessen's rule [1] and effect of far field radiation as

$$\Gamma_I = \frac{v_F}{\lambda_B} + \frac{v_F}{r}, \Gamma_R = \frac{2e^2\omega_0^2}{3\varepsilon_0 m^* c^3}, \quad (2)$$

where $\lambda_B$ is due to inelastic collisions with the particle surfaces and $v_F$ is the Fermi velocity, also $\omega_1$ is known to be derived using

$$\omega_1^2 = \frac{qe}{4\pi m^* \varepsilon_0 n^2 d^3}, \quad (3)$$

where $q$ is the magnitude of the oscillating charge, $m^*$ is the optical effective electron mass, $n$ is the reflective index of the hosting material, and $e$ is the electron charge.

Solving the (1) simultaneously for different values of $\omega$ yields $k$ and $\alpha$ as a function of $\omega$. We also need the dependency of the $\omega_0$ as a function of the radius of the nanospheres. The plasmon dispersion relation of a single nanoparticle in the case of spherical boundary conditions can be derived from the following complex dispersion equation [5]

$$\sqrt{\varepsilon_{in}}\xi_l'(k_{out}r)\varphi_l(k_{in}r) - \sqrt{\varepsilon_{out}}\xi_l(k_{out}r)\varphi_l'(k_{in}r) = 0. \quad (4)$$

Now, $\omega_0$ can be found from the solutions of the equations 4, in which the wave numbers $k_{in}$, and $k_{out}$ are

$$k_{in} = \frac{\omega_0}{c}\sqrt{\varepsilon_{in}(\omega_0)}, k_{out} = \frac{\omega_0}{c}\sqrt{\varepsilon_{out}(\omega_0)}. \quad (5)$$

In addition, $\varepsilon_{in}$ and $\varepsilon_{out}$ are dielectric functions of the investigated nanosphere and the dielectric environment respectively. Considering non-dispersive environment and metallic nanosphere, the dielectric function is

$$\varepsilon_{in}(\omega_0) = 1 - \frac{\omega_p^2}{\omega_0^2}, \varepsilon_{out}(\omega_0) = 1, \quad (6)$$

and $\varphi_l$ and $\xi_l$ can be expressed from Bessel and Hankel functions respectively as

$$\varphi_l(z) = z\sqrt{\frac{\pi}{2z}} J_{l+\frac{1}{2}}(z)$$
$$\xi_l(z) = z\sqrt{\frac{\pi}{2z}} H^{(1)}_{l+\frac{1}{2}}(z)$$
(7)

## III. RESULTS AND ANALYSIS

In Fig. 1, the results of the optical absorption measurements for the deposited DLC are plotted. Here, the effect of substrate absorption is eliminated using transfer matrix method [16]. The optical spectrum depicted in Fig. 1 is in agreement with the previously reported results for the CVD diamond [17]. It should be noted that the DLC is the background material for the presented network of copper nanoparticles, which its process of deposition is described in [18].

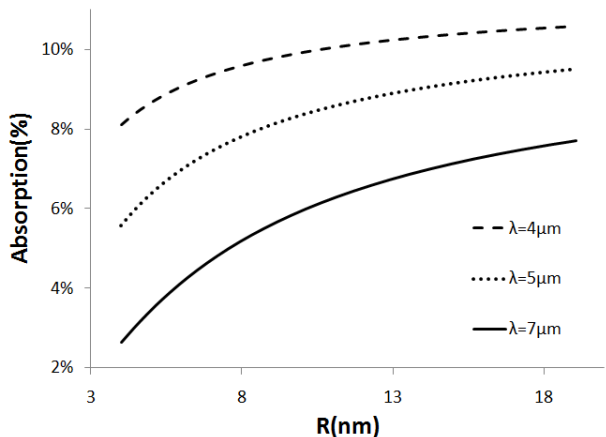

Fig. 4. Absorption spectrum for different particle radius.

Parallel chains of metallic nanoparticles are considered perpendicular to the surface of the deposited nanocomposites. By simultaneous solution of coupled equations (1) the absorption spectrum and dispersion relation of compound Hertzian chain are derived. The variation of the wavelength, $\lambda$ is limited to the range of 2.5 to 10μm. The radius of the nanoparticles is also changed from 4 to 18nm. The electronic relaxation frequency, $\Gamma_R$ and inelastic collisions, $\lambda_B$ are $8.2\times10^8$rad/s and 39nm respectively. The analytical and the experimental results are plotted for two different values of the copper content as shown in Fig. 1. The experimental results are published previously [14]. Here, we used the DLC absorption spectrum to eliminate its effect of the optical absorption of nanoparticle network. The analytically derived absorption decays to zero as the wavelength increases or frequency decreases. Analytical results seem to match the measurements considering particle center-to-center distances of 33nm and 41nm with constant particle radius of 15nm. These results are approximately in agreement with the previously published results of atomic force microscopy (AFM) images with regard to both of the samples that are fabricated in the same group [18]. The match between experiment and analytical curves at low wavelengths might result that the dominant absorption mechanism in this range is the plasmon effect of nanoparticle chain. However, the absorption predicted by this mechanism vanishes at higher wavelengths.

The coupled equations (1) are also solved for different values of nanoparticle interspacing and radius. Absorption spectrum for different nanoparticle interspacing, $d$, and constant radius of 15nm is depicted in Fig. 3. As observed in Fig. 3, the absorption decreases rapidly by increasing interspacing. The behavior of absorption versus interspacing is similar for different wavelengths; however, the absorption shifts downward for higher wavelengths. Absorption spectrum for different nanoparticle radius, $r$ and interspacing of 40nm is depicted in Fig. 4. As observed in Fig. 4, the absorption grows slowly by increasing radius, $r$, but in spite of previous curve this behavior is not the same for different wavelengths.

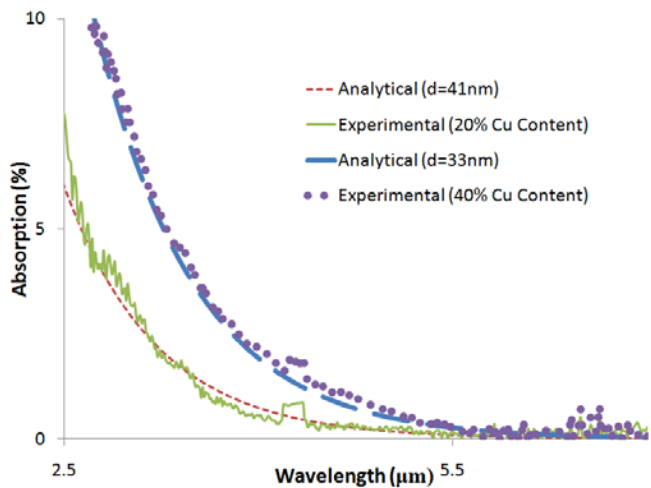

Fig. 2. Experimental and analytical absorption spectrum.

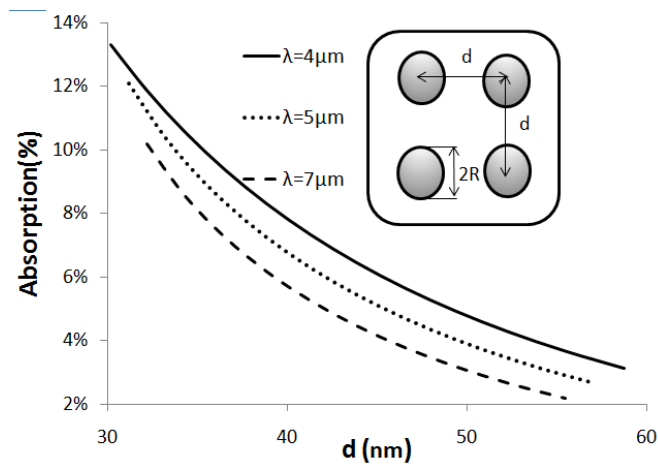

Fig. 3. Absorption spectrum for different particle interspacing.

## IV. Conclusion

Absorption mechanism of metallic nanoparticle network based on the plasmon wave solution of dipole chains has been investigated in this work. The nanoparticle composite is modeled as neighboring chains of coupled dipoles, in which each chain is considered parallel to electromagnetic radiation. We introduced the effects of damping factors like interaction with phonons, electrons, lattice defects, and radiation into the far field to explain the decay of the plasmon wave through the chain. Then the absorption coefficient is derived by solving the equation of motion of oscillating charges in the presence of dispersive and damping forces. For this calculation, plasmonic self-resonance frequency has been obtained through the solution of dispersion relation of a single nanoparticle by applying the spherical boundary conditions. Applying this approach to the network of Cu nanoparticles, the spectrum of the absorption versus wavelength in the range of micrometer is calculated analytically. We also investigated the effects of nanoparticle size and interspacing in the absorption spectrum. Analytical results are compared with the previously reported experimental data and good agreement is observed in considered wavelength range.